\documentstyle[aps,prd,preprint,mathrsfs]{revtex}
\def\bw{\bar\omega}
\def\ba{\bar\alpha}
\def\bb{\bar\beta}
\def\half{\frac{1}{2}}
\def\ts{\tilde s}
\def\sr{s_{R}}

\begin{document}
\tightenlines



\title{A nilpotent symmetry of quantum gauge theories}   

\author{Amitabha Lahiri}
\address{S. N. Bose National Centre for Basic Sciences, \\
Block JD, Sector III, Salt Lake, Calcutta 700 098, INDIA}
\address{amitabha@boson.bose.res.in}
\date{Final Version, August 23, 2001}

\maketitle

\begin{abstract}
For the Becchi-Rouet-Stora-Tyutin (BRST) invariant extended action
for any gauge theory, there exists another off-shell nilpotent
symmetry.  For linear gauges, it can be elevated to a symmetry of
the quantum theory and used in the construction of the quantum
effective action. Generalizations for nonlinear gauges and actions
with higher order ghost terms are also possible.
\end{abstract}
\bigskip



{\em Introduction:} Quantization of gauge theories requires
gauge-fixing, and for most gauges, the introduction of ghost
fields. The resulting theory is invariant, not under the gauge
symmetry itself, but under the Becchi-Rouet-Stora-Tyutin (BRST)
symmetry\cite{brs,tyu}. Nilpotence of the BRST transformation
allows it to be extended to a symmetry of the quantum theory at all
orders of the perturbation series, which allows order by order
cancellation of infinities by the introduction of appropriate
operators in the action. The quantum effective action is then the
most general function invariant under this symmetry as well as all
other known quantum symmetries of the theory.

In any useful gauge theory, gauge fields are coupled to many other
fields. For general gauge theories, several of these fields may
have the same Lorentz and gauge transformation properties. This
leads to an enormous number of possible terms in the quantum
effective action. If the theory is renormalizable, most of these
terms have to vanish, leaving only those which are identical with
the tree-level action, up to multiplicative constants. The BRST
symmetry, imposed through the Slavnov-Taylor operator, ensures this
stability for the physical ghost-free gauge-invariant part of the
action. The demonstration of stability of the gauge-fixing and
ghost terms requires auxiliary conditions.

In Landau-type gauges, the auxiliary conditions used are the ghost
equation, the antighost equation, and their commutators with the
Slavnov-Taylor operator\cite{piso}. In more general linear gauges
the antighost equation picks up a nonlinear breaking term and thus
loses its usefulness. This is particularly inconvenient for linear
interpolating ($R_\xi$ type) gauges. It is also inconvenient for
gauge theories involving several fields with similar group and/or
Lorentz transformation properties. For example, in theories
involving non-Abelian two-form fields, one finds auxiliary vector
fields and corresponding scalar ghosts. These can mix with the
usual vector or ghost fields. As a result the proof of stability of
general linear (including interpolating) gauges in such theories
can be quite long. For nonlinear gauges and for actions containing
terms of quartic or higher order in the ghost fields, proofs of
stability are even more complicated.

In this paper I show that the BRST-invariant extended action for
any gauge theory admits another, gauge fermion dependent, nilpotent
symmetry, which does not seem to have been noticed earlier. This
symmetry differs from the BRST symmetry only in its action on
trivial pairs. For some special kinds of action, for example those
which are quadratic in ghost fields, it becomes identical with the
BRST symmetry upon using equations of motion. However, off shell it
is always different from BRST, and can be used as an auxiliary
condition to uniquely determine the quantum effective action of a
gauge theory, including ghost and gauge-fixing terms. This symmetry
holds in general linear gauges in addition to BRST, so it is
particularly convenient for proving the uniqueness and stability of
the ghost and gauge fixing sector of gauge theories outside Landau
gauge, unlike the usual algebraic renormalization scheme.  Below, I
construct this symmetry, first when the theory has only fermionic
ghosts, and then for a theory with both fermionic and bosonic
ghosts. As a simple illustration I will apply this construction to
the example of Yang-Mills theory, but the real convenience of this
symmetry becomes apparent when it is applied to theories with
larger field content, such as theories of $p$-form gauge fields. A
generalization of the construction, somewhat similar to the well
known antifield construction (see \cite{Barnich:2000zw} for a
review) suggests itself for theories with higher order ghost terms,
and is discussed at the end of the paper.

The extended ghost sector of the tree-level quantum action of a
gauge theory can be written in the general form
\begin{eqnarray}
S^c_{ext} =   h^A f^A + \half \lambda h^A h^A +
\bw^A \Delta^A \,.
\label{1199.Scext}
\end{eqnarray}
The anticommuting antighosts $\bw^A$ and the corresponding
auxiliary fields $h^A$ form what are known as trivial pairs. Here
the index $A$ stands for the collection of all indices, $f^A = 0$
is the gauge-fixing condition, $\lambda$ is a constant gauge-fixing
parameter, and $\Delta^A$ is the BRST variation of the gauge-fixing
function, $\Delta^A = sf^A$. The sum over $A$ includes the
integration over space-time, and $f^A$ has been chosen to be
independent of $\bw^A$ and $h^A$. This part of the action remains
invariant if the trivial pair transform under BRST as
\begin{eqnarray}
s \bw^A = -h^A\,, \qquad s h^A = 0 \,,
\label{1199.brs}
\end{eqnarray}
and can be written as a BRST differential of a gauge-fixing
fermion $\Psi$,
\begin{eqnarray}
S^c_{ext} = s\left( - \bw^Af^A - \half\lambda \bw^Ah^A \right)
\equiv s\Psi \,.
\label{1199.sPsi}
\end{eqnarray}
On the other hand, I can rearrange $S^c_{ext}$ as
\begin{eqnarray}
S^c_{ext} &=& \half \lambda \left( h^A + {1\over \lambda}
f^A\right) \left( h^A + {1\over \lambda} f^A\right) - {1\over
2\lambda} f^A f^A + \bw^A \Delta^A \nonumber \\ 
&=& \half \lambda \left( \left(h^A + {2\over \lambda} f^A \right) -
{1\over \lambda} f^A\right) \left( \left( h^A + {2\over \lambda}
f^A\right) - {1\over \lambda} f^A\right) 
- {1\over 2\lambda} f^A f^A + \bw^A \Delta^A \nonumber \\
&=& h'^A f^A + \half \lambda h'^A h'^A + \bw^A \Delta^A \,,
\label{1199.newh}
\end{eqnarray}
where I have defined $h'^A = - h^A - \displaystyle{2\over \lambda}
f^A$.  Now $S^c_{ext}$ has the same functional form as before, but
in terms of a redefined auxiliary field $h'^A$. It follows that
$S^c_{ext}$ is invariant under a new set of transformations:
\begin{eqnarray}
\ts \bw^A = - h'^A &\Rightarrow& \ts\bw^A = h^A + {2\over\lambda}
f^A \,,\nonumber \\ 
\ts h'^A  = 0 &\Rightarrow& \ts h^A = - {2\over\lambda} \ts f^A \,,
\nonumber \\ 
\ts &=& s\qquad {\mathrm on\; all\; other\; fields}.
\label{1199.cstilde}
\end{eqnarray}
It follows that $\ts$ is nilpotent on all fields, $\ts^2 = 0$. It
should be emphasized that $\ts$ is a symmetry of the original
($s$-invariant) action itself, not some special feature of the
construction procedure.

When the extended sector corresponds to the gauge-fixing of an
anticommuting gauge field, as can happen for theories with
reducible gauge symmetries, the construction is slightly more
complicated, since the auxiliary fields now have odd ghost
number. Typically, the extended ghost sector in this case can be
written with anticommuting auxiliary fields $\ba^A$, $\alpha^A$ as 
\begin{eqnarray}
S^a_{ext} = \ba^A f'^A + \bar f'^A \alpha^A + \zeta \ba^A \alpha^A
+ \bb^A \Delta'^A \,.
\label{1199.Saext}
\end{eqnarray}
In this, $f'^A$ is the anticommuting gauge-fixing function,
$\Delta'^A = s f'^A$, and $\bb^A$ is the corresponding commuting
antighost. The term $\bar f'^A \alpha^A$ is a rearrangement of the
appropriate terms in $\bw^A \Delta^A$ which appear in $S^c_{ext}$
of Eq.~(\ref{1199.Scext}) for the usual gauge symmetries.  Such
terms are not affected by the redefinitions in
Eq.~(\ref{1199.newh}), so they will appear in
Eq.~(\ref{1199.Saext}). In addition to Eq.~(\ref{1199.brs}), the
BRST transformations on the extended sector now include $s\bb^A =
\ba^A, s\ba^A = s\alpha^A = 0$, and $s(S^c_{ext} + S^a_{ext}) = 0$,
although $S^c_{ext}$ and $S^a_{ext}$ are not separately
BRST-invariant.

Just as in the case with commuting auxiliary fields, the terms in
$S^a_{ext}$ can be rearranged,
\begin{eqnarray}
S^a_{ext} &=& \zeta \left( \ba^A + {1\over \zeta} \bar f'^A \right)
\left( \alpha^A + {1\over \zeta}  f'^A \right) -  {1\over \zeta}
\bar f'^A f^A + \bb^A \Delta'^A \, \nonumber \\
&=& \zeta \left( \left( \ba^A + {2\over \zeta} \bar f'^A \right) -
{1\over \zeta} \bar f'^A \right)  \left( \left( \alpha^A + {2\over
\zeta} f'^A \right) - {1\over \zeta}  f'^A \right) -  {1\over
\zeta} \bar f'^A f^A + \bb^A \Delta'^A \, \nonumber \\
&=& \zeta \ba'^A \alpha'^A + \ba'^A f'^A + \bar f'^A \alpha'^A +
\bb^A \Delta'^A \,.
\label{1199.newalpha}
\end{eqnarray}
where I have now defined $\ba'^A = - \left( \displaystyle \ba^A +
{2\over \zeta}\bar f'^A \right)$ and $\alpha'^A = - \left(\alpha^A
+ \displaystyle {2\over \zeta} f'^A \right)$.  As before, a new set
of BRST transformations can be defined for $S^a_{ext}$,
\begin{eqnarray}
\ts \bb^A &=& \ba'^A = - \left( \ba^A + {2\over \zeta} \bar f'^A
\right) \,, \nonumber \\
\ts \ba'^A &=& 0 \Rightarrow \ts \ba^A = - {2\over \zeta} \ts \bar f'^A
\,, \nonumber \\
\ts \alpha'^A &=& 0 \Rightarrow \ts \alpha^A = - {2\over \zeta} \ts f'^A
\,, \nonumber \\
\ts &=& s\qquad {\mathrm on\; all\; other\; fields}\,.
\label{1199.astilde}
\end{eqnarray}
Since $\alpha^A$ was the result of BRST variation of some field,
$\alpha'^A$ has to be the variation under $\ts$ of the same field,
and $\ts \bar f'^A$ must be calculated according to the rules of
Eq.~(\ref{1199.cstilde}). In addition, the action of $\ts$ must be
the same as that of $s$ for the fields contained in $f'^A$. Then
$\ts^2 = 0$ on all fields, and $\ts(S^c_{ext} + S^a_{ext}) = 0$.

\bigskip
{\em Example:}
Let me consider a concrete example, and construct this symmetry for
Yang-Mills theory in an arbitrary (linear or nonlinear)
gauge-fixing function $f^a$. The tree-level quantum action is in
this case
\begin{equation}
S = \int d^4x\, \bigg(- {1\over 4}F_{\mu\nu}^aF^{a\mu\nu} + h^a f^a
+ \bw^a \Delta^a + \half\xi h^a h^a \bigg) \,,
\label{1199.YMtree}
\end{equation}
where $a$ is the gauge index.
This is invariant under the BRST transformations
\begin{eqnarray}
s A^a_\mu &=& \partial_\mu \omega^a + g f^{abc}A^b_\mu \omega^c\,,\qquad
s \bw^a = - h^a \,, \qquad \nonumber \\ 
s \omega^a &=& - {1\over 2}
gf^{abc}\omega^b\omega^c \,, \qquad 
s h^a = 0\,. \label{1199.brst}
\end{eqnarray}
Following the rules of Eq.~(\ref{1199.cstilde}), I obtain
\begin{eqnarray}
\ts \bw^a &=& h^a + {2\over \xi}\, f^a\,, \qquad \ts h^a
= -{2\over \xi} \Delta^a \,, \qquad\nonumber \\
\ts &=& s \mbox{ on all other fields.} \label{1199.stilde} 
\end{eqnarray}
By construction $\ts$ is a symmetry of the action, $\ts S = 0$, and
nilpotent, $\ts^2 = 0$. It is straightforward to check these two
properties explicitly for this example of Yang-Mills theory.

Any symmetry is a useful property of a theory, just how useful
depends on both the symmetry and the theory. Let me show here how
this symmetry can be used jointly with BRST to ease calculations.
The quantum effective action $\Gamma[\chi, K]$ defined in the
presence of background c-number sources $K^A$ for the BRST
variations $F^A = s\chi^A$ obeys the Zinn-Justin equation \cite{ZJ}
$(\Gamma, \Gamma) = 0$, where $(\cdot, \cdot)$ is the antibracket
in terms of $\chi^A$ and their sources for BRST variations, $K^A$.
Note that $\Gamma[\chi, K]$ does not contain the sources for the
BRST variations of auxiliary fields of the type $h^A,
\alpha^A,\ba^A$, etc. which are BRST invariant. Also note that
$\Gamma_{N, \infty}$, which is the infinite part of the $N$-loop
contribution to $\Gamma$, does not contain the sources for the BRST
variations of antighosts of the type $\bw^A$, because their BRST
variations are linear in the fields \cite{weinqft}.

For most physically interesting cases the effective action is at
most linear in the remaining $K^A$ on dimensional
grounds. This is the case for pure Yang-Mills fields, as well as
several theories with Yang-Mills fields coupled to various other
fields, in four dimensions. For these theories, the Zinn-Justin
equation reduces to the statement that for infinitesimal
$\epsilon$, $S_R + \epsilon \Gamma_{N, \infty}$ is invariant under
the quantum BRST symmetry $\sr$, which is just the most general
nilpotent symmetry built out of the fields in the theory, and which
reduces to the original BRST symmetry at tree-level \cite{weinqft}.

Let me look at this class of theories, viz., those for which
$\Gamma[\chi, K]$ has been shown to be at most linear in the $K^A$.
Let me also assume that the quantum BRST transformation $\sr$ has
been found by solving the Zinn-Justin equation.  In order to see
the effect of the gauge-dependent symmetry $\ts$ on the quantum
theory, I take the same effective action $\Gamma[\chi, K]$ with the
same sources.  Of course $\ts$ is a gauge-dependent symmetry,
nonetheless it can be elevated to a symmetry of the quantum
effective action if the gauge-fixing functions are linear in the
fields. I shall denote the minimal fields by $\phi^A$ and
non-minimal fields by $\lambda^A$.  Then $\ts\phi^A = s\phi^A$, and
consequently $\ts s\phi^A = 0$. The application of $\ts$ on the
partition function gives (since the tree-level action $S$ is
invariant under $\ts$),
%
\begin{equation}
\langle F^A \rangle {\delta_L\Gamma[\chi, K]\over
\delta \phi^A}+ \langle \ts\lambda^A \rangle {\delta_L\Gamma[\chi,
K]\over \delta \lambda^A} + \langle \ts s\lambda^A \rangle
K^A[\lambda] = 0\,.
\label{1199.ts}
\end{equation}
%
Here $\langle\;\rangle$ denotes the quantum average in the presence
of sources, specified such that the quantum average of a field is
the field itself \cite{weinqft}. So far the gauge could be
arbitrary. For the special case where the gauge-fixing functions
are assumed to be linear in the fields, $\ts\lambda^A$ as defined
in Eq.~(\ref{1199.cstilde}) and Eq.~(\ref{1199.astilde}) is
either linear in the fields or equals the BRST variation of some
linear function of the fields. Either way, $\langle \ts \lambda^A
\rangle$ is known explicitly. In addition, the effective action
does not contain the sources for BRST variations of $(h^A,\alpha^A,
\ba^A)$ etc. and only $S_R$ contains the sources for the BRST
variations of $(\bw^A, \bb^A)$ etc. Then I can read off from
Eq.~(\ref{1199.ts})\ that $S_R + \epsilon\Gamma_{N, \infty}$ is
invariant under $\ts_R$, which is just $\ts$ as calculated in terms
of the quantum BRST transformation $\sr$.

Going back to the example of Yang-Mills theory in four dimensions,
I obtain directly from Eq.~(\ref{1199.ts}) that in a linear gauge
the quantum symmetry corresponding to $\ts$ is given just by
Eq.~(\ref{1199.stilde}) with $s$ and $\ts$ replaced by $\sr$ and
$\ts_R$, respectively, where $\sr$ is the usual quantum BRST
transformation for Yang-Mills fields \cite{ZJ,weinqft}.  The ghost
sector of the general effective action can now be obtained through
an extremely short calculation. Let me define $\sr' = \half(\sr -
\tilde\sr)$.  Then
\begin{eqnarray}
\sr'\bw^a &=&  -\left( h^a + {1\over\xi}f^a \right)\,,
\qquad \sr' h^a =  {1\over \xi}\Delta^a_R\,, \nonumber \\
\qquad \sr' &=& 0 \mbox { on all other fields}.
\end{eqnarray}
Since Yang-Mills theory is power-counting renormalizable in four
dimensions, the infinite part of the $N$-loop quantum effective
action, after infinities up to $N-1$ loops have been absorbed into
counterterms, is an integrated local functional of mass dimension
four. So on dimensional grounds and because the effective action
must have zero ghost number, it can be at most quadratic in the
trivial pair $\lambda^A \equiv (\bw^a, h^a)$ \cite{weinqft}. So I
can write
\begin{equation}
\Gamma = S_C + \lambda^A X^A + \lambda^A\lambda^B X^{AB}\,,
\label{1199.Gamma}
\end{equation}
where $S_C$ does not contain any ghost or auxiliary field, and
$X^A$ and $X^{AB}$ do not contain any of the $\lambda^A$.  Then the
coefficients of different powers of the $\lambda^A$ in the equation
$\sr'\Gamma = 0$ must vanish. In particular, the terms quadratic or
linear in $\lambda^A$ give
\begin{eqnarray}
X^{ab}_{\bw\bw} = 
X^{ab}_{\bw h} &=& 0\,, \nonumber \\ 
- X^a_{\bw} + {2\over\xi}\Delta^b_R  X^{ab}_{hh} &=& 0,
\end{eqnarray}
while the terms independent of $\lambda^A$ in $\sr'\Gamma = 0$ give
\begin{equation}
- f^a X^a_{\bw} + \Delta^a_R X^a_h = 0\,.
\label{1199.nolambda}
\end{equation}

In addition, antighosts and auxiliary fields transform among
themselves under BRST, so I can also consider the coefficients of
$\lambda^A$ in $\sr\Gamma = 0$ to obtain some independent
equations,
\begin{equation}
\sr X^a_{\bw} = 0 = \sr X^{ab}_{hh}, \qquad X^a_{\bw} = \sr
X^a_h\,. 
\label{1199.sRlambda}
\end{equation}
Here the function $ X^{ab}_{hh}$ is symmetric and has vanishing
mass dimension and ghost number.  It follows from the above
equation that $X^{ab}_{hh}$ is purely numerical, and because we are
dealing with the SU(N) algebra, and $X^{ab}_{hh}$ is clearly
symmetric in $(a,b)$, it must be proportional to $\delta^{ab}$. Then
\begin{equation}
X^{ab}_{hh} = {\xi Z_\omega\over 2}\delta^{ab}, \quad X^a_{\bw} =
Z_\omega \Delta^a_R \quad \mbox{and}\quad  X^a_h = Z_\omega f^a\,,
\end{equation}
for some constant $ Z_\omega$. (The last equation follows from
combining Eq.s~(\ref{1199.sRlambda}) and (\ref{1199.nolambda}).)
Therefore the quantum effective action takes the form
\begin{equation}
\Gamma = S_C + Z_\omega\bw^a\Delta^a_R + Z_\omega h^a f^a + 
{\xi \over 2}Z_\omega h^a h^a \,,
\label{1199.qea}
\end{equation}
where $S_C$ is the most general ghost-free polynomial of dimension
four symmetric under $\sr$ and all linear symmetries of the
classical theory. Note that I did not need to assume any specific
gauge-fixing function, only that it is linear.

It is known that the problem of stability of the ghost (and
gauge-fixing) sector of gauge theories can be solved by using the
ghost and antighost equations as auxiliary conditions in the usual
algebraic renormalization scheme~\cite{piso}. However, those
equations are in their most useful form in the Landau gauge $\xi =
0$, while the symmetry $\ts$ is defined for a non-zero $\xi$ and
cannot even be constructed directly for $\xi = 0$.  (Of course, the
$\xi \to 0$ limit can be taken after the effective action has been
found.)  Therefore the symmetry $\ts$ is not a reformulation of the
usual auxiliary conditions. In particular, the use of $\ts$ as an
auxiliary condition in the algebraic renormalization scheme, as
opposed to the ghost and antighost equations, can be thought of as
being complementary to those auxiliary conditions. This symmetry is
especially useful in dealing with Yang-Mills type theories with a
large number of fields, for which various different interpolating
linear gauges are allowed.  Examples are theories with $p$-form
fields, as in the first order formulation of Yang-Mills theory
\cite{Fucito:1997sq} or the topological mass generation mechanism.
The technique described here provides a straightforward way of
verifying the uniqueness of the gauge-fixing and ghost sector of
such theories, as has been done in \cite{renorm}.

\bigskip
{\em Generalizations:} The calculations for the example were done
assuming that the gauge condition is linear in the fields. This was
mainly for convenience --- just as for the usual BRST symmetry,
results in linear gauges are easier to calculate and interpret. But
even in nonlinear gauges, or for actions with quartic ghost terms,
there is a corresponding nilpotent symmetry. Let me show the
construction for Yang-Mills theory in four dimensions,
generalizations to many other cases being fairly simple. First, the
gauge-fixing fermion of Eq.~(\ref{1199.sPsi}) is generalized to
include terms quadratic in the antighost, so that
\begin{eqnarray}
S^c_{ext} &=&  s\Psi = s( -\bw^af_0^a - \half\xi\bw^ah^a -
\half\bw^a\bw^bf_1^{ab} )\,\nonumber\\
&=& \bw^a\Delta_0^a + h^af_0^a + \half\xi h^ah^a + h^a\bw^bf_1^{ab}
-\half\bw^a\bw^b\Delta_1^{ab}\,,
\label{1199.quartic}
\end{eqnarray}
where $f_0^a$ and $f_1^{ab}$ do not contain $\bw^a$ or $h^a$, but
are arbitrary otherwise, and $\Delta_0^a = sf_0^a$ and
$\Delta_1^{ab} = sf_1^{ab}$.  For Yang-Mills theory in four
dimensions, $\Psi$ must be of dimension three or less, so there are
no further terms. Now I can `complete the square' as before, and
write
\begin{eqnarray}
S^c_{ext} &=& \half\xi\left(h^a + {1\over\xi}f^a \right) \left(h^a
+ {1\over\xi}f^a \right) - {1\over 2\xi}f^af^a + \bw^a\Delta_0^a -
\half\bw^a\bw^b\Delta_1^{ab}\,,
\end{eqnarray}
where $f^a =f_0^a + \bw^bf_1^{ab}$. Then as before I can define
$\displaystyle h'^a = -\left(h^a + {2\over\xi}f^a\right)$ and write
\begin{eqnarray}
S^c_{ext} = \bw^a\Delta_0^a + h'^af_0^a + \half\xi h'^ah'^a +
h'^a\bw^bf_1^{ab} -\half\bw^a\bw^b\Delta_1^{ab}\,,
\end{eqnarray}
which has the same functional form as Eq.~(\ref{1199.quartic}),
but with $h^a$ replaced by $h'^a$. So the new symmetry
transformations are 
\begin{eqnarray}
\ts\bw^a  &=& h^a + {2\over\xi}f_0^a +
{2\over\xi}\bw^bf_1^{ab}\,,\nonumber\\ 
\ts h^a &=& - {2\over\xi}\Delta_0^a -
{2\over\xi}h^bf_1^{ab} - {4\over\xi^2} f_0^bf_1^{ab} -
{4\over\xi^2} \bw^cf_1^{bc}f_1^{ab} +
{2\over\xi}\bw^b\Delta_1^{ab}\,,\nonumber\\
\ts &=& s\qquad \mbox { on all other fields}.
\end{eqnarray}
Again, this is a symmetry of the action $S^c_{ext}$, and therefore
a symmetry of the full action including the gauge-invariant terms.
Note that $\ts$ is again nilpotent by construction, $\ts^2 = 0$.

For the examples given so far, including the last one, $\ts$
differs from $s$ by a `trivial symmetry' \cite{hentei},
\begin{eqnarray}
\ts - s = \eta^{AB}{\delta S\over\delta\chi^A}{\delta\over
\delta\chi^B}\,, 
\end{eqnarray}
where in this case $\chi^A$ is restricted to run over $(\bw^A,
h^A)$ and $\eta^{AB}$ is graded antisymmetric in $(A, B)$.  What
makes $\ts$ special is the fact that it is nilpotent, since adding
a trivial symmetry to BRST does not make an off-shell nilpotent
symmetry in general. On the other hand, for general BRST-invariant
actions, the two symmetries $s$ and $\ts$ need not be related by a
trivial symmetry. For general actions, i.e., those which may
include higher order ghost terms, the construction of $\ts$ can be
generalized to give a nilpotent symmetry.  To see how that can be
done, note that for the examples above, $h'^A =
\delta\Psi/\delta\bw^A$ up to a constant coefficient, as if $h'^A$
were the antifield of $\bw^A$. It is worth emphasizing that $h'^A$
is not the antifield for $\bw^A$. But this similarity suggests a
generalization of the previous constructions in the following way.

Given a gauge invariant action $S_0$, let the ghost fields be
defined as usual, and introduce a trivial pair $(\bw^A, h^A)$ for
each generator, with the BRST transformation law $s\bw^A = -h^A,\,
sh^A = 0$. The gauge-fixing fermion $\Psi$ is then constructed as
some functional of ghost number\, $-1$, subject to any other known
symmetry or dimensional restriction. The ghost sector of the action
is then $s\Psi$, so that the total action $S_0 + s\Psi$ is
BRST-invariant. Now let a new BRST transformation $\ts$ be defined
as $\ts\bw^A = -h'^A,\, \ts h'^A = 0$, where $h'^A =
\delta\Psi/\delta\bw^A$, and $\ts = s$ on all other fields.  A new
gauge fixing fermion $\Psi'$ is then constructed by replacing $h^A$
by $h'^A$ in $\Psi$, i.e., $\Psi' = \Psi[h^A \to h'^A]$, and a new
ghost action is constructed as $\ts\Psi'$. The (new) total action
$S_0 + \ts\Psi'$ is then invariant under $\ts$.

If $\Psi$ is chosen to be the most general gauge fixing fermion,
$s\Psi$ would be the most general $s$-exact functional of vanishing
ghost number.  But now there are two actions, one constructed with
$s\Psi$, and the other with $\ts\Psi'$, and these two need not be
equal when written in terms of the same $h^A$. For the situations
where (as in all the examples above) $\ts\Psi[h^A \to h'^A] =
s\Psi$ up to a finite number of irrelevant constants, the total
action is invariant under two different off-shell nilpotent
symmetries $s$ and $\ts$. This can be an immensely useful property
for proving the uniqueness of the ghost action for complicated
theories. In addition, since $\ts$ differs from BRST
transformations only by its action on the trivial pair, it has the
same cohomology as the BRST transformation itself
\cite{Barnich:2000zw}.  So there is no additional complication in
calculating the structure of anomalies in the theory, which is
determined fully by the BRST cohomology.

In summary, any BRST invariant action in linear or nonlinear gauge
has another off-shell, nilpotent symmetry with the same cohomology
as the BRST transformation. If the action contains up to quartic
ghost terms, it is always symmetric under both BRST and this
transformation. If it contains higher order ghost terms, one can
construct another action which is symmetric under the new BRST
transformation, and whose gauge-invariant component agrees with
that of the original action. If the ghost sectors of the two
actions agree as well, both transformations leave it invariant.
This can simplify calculations of the counterterms, especially when
the gauge-fixing term is linear in the fields.

\bigskip
{\em Acknowledgement:} It is a pleasure to thank M.~Henneaux for a
helpful comment.


\end{document}